\documentstyle[preprint,aps]{revtex}
\input epsf
\draft
\tighten
\begin{document}
\preprint{TUM/T39-98-13} 

\title{Hadronic properties of the $S_{11}(1535)$ studied by electroproduction
off the deuteron}

\author{Leonid Frankfurt\thanks{St.Petersburg Nuclear Physics
      Institute, Gatchina, Russia}}
\address{School of Physics and Astronomy, Tel Aviv University, 
Tel Aviv 69978, Israel }

\author{Mikkel Johnson}
\address{Los Alamos National Laboratory, Los Alamos, New Mexico 87545, USA}

\author{Misak Sargsian \thanks{Yerevan Physics Institute, Yerevan, Armenia}
\thanks{after 1-May-1998 at University of Washington, Seattle, WA, USA} 
and Wolfram Weise} 
\address{Physik Department, Technische Universit\"at M\"unchen, 
D-85747 Garching, Germany}
 
\author{Mark Strikman\thanks{St.Petersburg Nuclear Physics
      Institute, Gatchina, Russia}}
\address{Physics Department, Pennsylvania State University, University
Park, PA 16802, USA}

\author{}

\date{\today{}}
\maketitle

\begin{abstract}
Properties of excited baryonic states are investigated in the context of
electroproduction of baryon resonances off the deuteron.  In particular, 
the hadronic radii and the compositeness of baryon resonances are studied 
for kinematic situations in which their hadronic reinteraction is the 
dominant contribution. Specifically,  we study the reaction $d(e,e'S_{11})N$ 
at $Q^2\ge 1~GeV^2$ for kinematics in which the produced hadronic state 
reinteracts predominantly with the spectator nucleon.  A comparison of 
constituent quark model and effective chiral Lagrangian calculations of 
the $S_{11}$ shows substantial sensitivity to the structure of the produced 
resonance.
\end{abstract}

\section{Introduction}

Characterizing the  structure of hadrons in the nonperturbative region
of QCD is  a fundamental issue for  understanding nuclear dynamics. One of 
the key windows for studies of low-energy QCD is the investigation of 
baryonic  resonance properties.  Both hadronic and electromagnetic 
interactions off a free nucleon with excitation of a particular resonance 
state have been used to carry out such a investigations (for a recent review 
of this subject see Refs.{\cite{Burkert,Stoler}).

In this paper, we address the problem of the investigation of baryonic 
resonances by studying their electroproduction from a deuteron i.e. 
$d(e,e'R)N$.  
Emphasis is to be given to kinematics in which the dominant 
contribution arises when a resonance produced on one of the nucleons 
(either the neutron or the proton) undergoes a soft elastic 
rescattering off the other (spectator or recoil) nucleon.  
As we will argue, such experiments then offer the possibility of 
determining the specific properties of the baryonic  resonance 
that are not possible to ascertain by studying 
production off  a single  nucleon.  

One such property is the hadronic radius of the resonance, a quantity 
that is crucial for understanding  the dynamics governing its composite 
hadronic structure.  Indeed, in a quantum mechanical potential picture, 
the {\em virial theorem} shows that the averaged Hamiltonian $(H)$ of the 
system is determined by the 
potential $(V)$ and its gradient $(\vec \nabla V)$:
\begin{equation}
 <\psi|H|\psi> = <\psi|V|\psi> + {1\over 2}
<\psi|\vec r \cdot \vec \nabla V(r)|\psi>.
\label{vir}
\end{equation}
{}From the above equation,  e.g. for the Coulomb potential, one obtains 
$<{1\over r}> = {mZ\alpha\over n^2}$, while for harmonic oscillator 
potential $<r^2> = {(n + {1\over 2})\over mw}$. Here 
$n$ is the principal quantum number.  The above examples illustrate that at
least for certain types of potentials, excited states have
larger radii than ground states.  Moreover, knowing the dependence of the 
 radius on the quantum number of the excitation may allow one to determine the 
interaction.

The issue of the radius of excited hadronic states also 
crucial for understanding the duality between the quark-gluon and the 
hadronic descriptions of the strongly interacting system. 
Indeed, several experimental observations, such as the $A$ dependence of 
the coherent photoproduction of $J/\Psi$ mesons from nuclei
\cite{FNAL86}, the 
energy dependence of the diffractive electroproduction of 
vector mesons with coherent nuclear recoil
\cite{FNAL94,HERA}, and coherent pion 
diffraction into two jets\cite{FNAL97}, indicate a 
reasonably  broad distribution (fluctuation)
of the interaction cross section for color singlet 
objects (possibly indicating color transparency and color opacity phenomena; 
see Ref.\cite{FMS94} for details).
To saturate  the sum rule for the distribution of the 
cross section, one should assume a significant probability for the cross 
section to be larger than average. 
One of the ideas for realizing such large cross sections 
is to adopt  larger sizes for hadronic resonances.

Another property of interest is the compositeness of the 
produced state -  whether it be a single baryonic resonance 
or the superposition of multichannel $meson-baryon$ components with 
a strong attraction that makes a resonance-like quasi-bound system.
{}From the point of view of the production of excited hadronic 
states from an isolated nucleon, it is difficult to identify a signature
to distinguish a single 
baryonic resonance from a multichannel $meson-baryon$ system 
when some of the channels contain a strong attraction, thus 
imitating the resonance-like mass distribution.

\vspace{0.4cm}

However if the nucleon is inside the nucleus, then the 
resonance or multichannel nature of the excited hadronic 
state would provide  a qualitatively different picture for 
hadronic reinteractions with the spectator nucleons of the nucleus.
This is the main idea which we are going to exploit in studying
properties of the excited hadronic states.

Specifically we will consider the electroproduction of the $S_{11}(1535)$ 
resonance in the $d(e,e'S_{11})n$ reaction, where the neutron will be 
detected as a spectator. The choice of the  $S_{11}(1535)$ is suggested 
by two important characteristics of this resonance:
first, the $S_{11}(1535)$
has a large cross section for the electromagnetic $NN^*$ transition and a 
weak $NN\rightarrow NN^*$ transition; and, second, the $S_{11}(1535)$ has a 
large $N\eta$ branching ratio (up to 55\%), which makes it experimentally 
easily detectable.
To gain a feeling for the sensitivity of the $d(e,e'S_{11})n$
reaction to the hadronic structure of the $S_{11}(1535)$, we will 
obtain predictions from two basically different approaches.
In the first case, the $S_{11}(1535)$ is represented as an excited 
baryonic state whose structure is described in terms of quark 
constituents.  For the second, the $S_{11}(1535)$
represents a strongly enhanced structure in the amplitude  
of a four-channel meson-baryon system with total 
isospin ${1\over 2}$.

Within the constituent quark model (CQM) classification, 
the $S_{11}(1535)$  belongs  to the $[70,1^-]_1$ 
supermultiplet and represents an  $L=1$ radial excitation of 
the nucleon (see e.g. Ref.\cite{Burkert}).  In a typical constituent-quark 
model with a harmonic oscillator ansatz for the interacting potential 
(e.g, \cite{size1,size2}), the larger radial extension of the 
$L=1$ wave function means that the distribution of quarks in the 
$S_{11}$ should be more spread out than the one for the nucleon.

A second approach is based on the framework developed in 
Refs.\cite{KSW,KWW,KSWU}, where effective potentials for the 
interaction of pseudoscalar Goldstone bosons ($\pi, \ K, \ \eta$) 
with octet baryons ($N, \ \Lambda, \ \Sigma, \Xi$)  have been derived 
from an $SU(3)$ effective  chiral Lagrangian (ECL), with QCD chiral 
symmetry breaking due to non-vanishing up, down and strange quark masses.  
While solving a coupled-channel Schr\"odinger equation with the above 
mentioned potentials for the four-channel system of $\pi N$, $\eta N$, 
$K\Lambda$, and $K\Sigma$ states with total isospin ${1\over 2}$, it was 
found in Refs.\cite{KSW,KWW} that the properties of the $S_{11}(1535)$ can 
be well-described as an $S$-wave superposition of these four states. 
Moreover, the strong $S$-wave attractive interaction of the $K\Sigma$ system 
can build a  quasi-bound state with the characteristics of the $S_{11}(1535)$. 
This approach naturally solves the problem of the large branching ratio of 
$S_{11}(1535)$ decay to $\eta$N, showing it to be a consequence of the 
strong coupling of the $\pi N$ and $\eta N$ channels to the $K\Sigma$ channel 
\cite{KSW,KWW,KSWU}. 
The fact that the properties of the $S_{11}(1535)$ do hardly change 
in nuclei, as observed in the nuclear photoproduction of $\eta$ 
meson\cite{RL}, is consistent with the ECL picture.

The CQM and ECL approaches described above reproduce fairly well the main 
features of the $S_{11}(1535)$ resonance, but it seems that reactions 
involving only one nucleon can not distinguish between these two models.  
However, scattering on the deuteron provides the possibility of a different 
reaction mechanism, namely the soft rescattering of the 
$S_{11}$ on the spectator nucleon.  Thus, if one selects kinematics in which 
the dominant contribution to $d(e,e'S_{11})n$ arises when the resonance 
produced from the proton rescatters on the spectator neutron, one will 
substantially increase the sensitivity of the reaction to the hadronic 
properties of the $S_{11}$ and possibly provide a means to distinguish the
two approaches.

As an independent development, in Refs.\cite{FMGSS} the quasielastic 
$d(e,e'N)N$ reaction was calculated within the generalized eikonal 
approximation~(GEA). It was demonstrated that one can indeed identify 
specific kinematics for which the dominant contribution comes from the 
rescattering of the knocked-out nucleon on the spectator nucleon. Such 
a contribution is provided at large transverse momenta  of the spectator 
nucleon with respect of the momenta of virtual photon. Calculations in
Ref.\cite{FMGSS} demonstrate that starting at $Q^2\ge 1~GeV^2$ the 
eikonal approximation is well justified: in its low-energy limit, with 
a transfered energy of $0.5~GeV$, the GEA agrees (within 5-10\%) with the 
results of the calculations of Ref.\cite{Aren}. In Ref.\cite{Aren}, the 
eikonal approximation was not used, but the contributions of a large 
number of partial-waves (up to eight) were summed. 

\vspace{0.4cm}

In this paper, we will incorporate the GEA method into the CQM and ECL 
frameworks for describing the $S_{11}$ and calculate the $d(e,e'S_{11})n$ 
reaction at $Q^2\ge 1~GeV^2$.  The results of the CQM will demonstrate that 
for kinematics where the rescattering contribution to the $d(e,e'S_{11})n$ 
reaction is dominant there is a substantial sensitivity to the hadronic 
size of the final state produced. Significantly different predictions are 
made within the CQM as compared to the ECL approximation.

We first (in Section II) set up the kinematics and calculational procedure 
for $d(e,e'S_{11})n$ within the generalized eikonal approximation. 
The structure of $S_{11}(1535)$ will then be incorporated using the 
constituent quark model and the chiral dynamics approach. 
In Section III, the results of our numerical estimates will be discussed.  
In Section IV, we will discuss a class of resonances that may be investigated 
in a similar way.  Finally, in Section V we will summarize the paper.

\section{Setting up the kinematics and cross section}  

\subsection{Kinematics}
As a first step, to establish the relevant kinematics, we consider the 
special situation for which the $S_{11}(1535)$ resonance is produced
off a nucleon with small Fermi momentum. One way to fix the kinematics 
corresponding to the production of the $S_{11}$ resonance off a quasifree 
nucleon (almost at rest) is to choose $x\equiv {Q^2\over 2mq_0}$ to be
\begin{equation}
x = 1 - {m_R^2 - m^2\over Q^2 + m_R^2 - m^2} .
\label{xbj}
\end{equation}
Here $q_0$ and $-Q^2$ are the  energy and four-momentum  squared of the
virtual 
photon, and $m$ and $m_R$ are the masses of the nucleon and the $S_{11}$ 
resonance, respectively. We define also the mass $W$ of the produced 
hadronic state as:
\begin{equation}
W^2 = (q + m_d - p_s)^2 , 
\label{W2}
\end{equation}
where $q\equiv (q_0, {\bf q})$  and $p_s\equiv (E_s, {\bf p}_s)$ are 
the four-momenta of the virtual photon and spectator nucleon respectively.

To  ensure that the process occurs off a nucleon with small Fermi 
momentum, we shall also require that the light-cone momentum of the recoil 
nucleon (fraction of the deuteron momentum carried by spectator nucleon) 
be near one \cite{FMGSS}, i.e.
\begin{equation}
{E_s - p_{sz}\over m} \approx  1.
\label{alpha}
\end{equation}
The $z$ axis is defined by the direction of virtual
photon ${\bf q}$. Note, however, that we will require that the transverse 
momentum of the spectator be $p_{st}\lesssim 0.4~GeV/c$ to ensure that
the dominant contribution arises from the rescattering diagram.
Using this restriction on the Fermi momentum, one can also neglect any 
contribution from the $N^*$ component of the deuteron ground state wave 
function and end up with the set of diagrams presented in Fig.1.

\subsection{ Cross Section of the Reaction}

In general, the differential cross section of the $d(e,e'R)n$ reaction, 
where the momentum of the  scattered electron and spectator nucleon are 
measured in the final state, can be represented as follows:
\begin{equation}
{d^6 \sigma\over dE'_e d\Omega_e d^3 p_s} = 
{2 \alpha\over Q^4} {E'_e\over E_e} \eta_{\mu\nu} w^{\mu,\nu}
\label{crs1}
\end{equation}
where $\alpha={1\over 137}$, the
four-momenta of the incoming and scattered electrons 
are $k_{e}^{\mu}=(E_e,{\bf k}_e)$ and $k_{e}^{\mu'}=(E'_e,{\bf k'}_e)$
respectively, 
and $\eta_{\mu\nu}= {1\over 2}Tr(\hat k'_e \gamma^{\mu} \hat k_e \gamma^{\nu})$
is the leptonic tensor. The hadronic tensor can be expressed through the 
electromagnetic transition amplitude of the deuteron $F^{\mu}$ as follows: 
\begin{equation}
w^{\mu,\nu} = \sum_{s_i}\limits^{-}\sum_{s_f} F^{\mu}F^{\nu\dag},
\label{hdt}
\end{equation} 
where we average over the initial and sum over the final spin states.
It is convenient to express the cross section 
in eq.(\ref{crs1}) through the four invariant functions 
$\sigma_T$, $\sigma_T$, $\sigma_{TT}$ and $\sigma_{TL}$ as follows:
\begin{equation}
{d^6 \sigma\over dE'_e d\Omega_e d^3 p_s} = {\alpha\over 2 \pi^2} 
{E'_e K\over Q^2 E_e (1-\epsilon)} 
\left\{\sigma_T + \epsilon\sigma_L - \epsilon cos(2\phi)\sigma_{TT} + 
\sqrt{\epsilon(\epsilon+1)}cos(\phi)\sigma_{TL}\right\}
\label{crs2}
\end{equation}
where $\epsilon = [1 + 2tan^2({\theta_e\over 2}){{\bf q}^2\over Q2}]^{-1}$ and
$\phi$ defines the azimuthal angle between the $(k_e,k'_e)$  and 
$(k_e,p_s)$ planes. The four invariant structure functions defined as:
\begin{eqnarray}
\sigma_T & = & {4\pi^2\alpha\over K}{w^{x,x}+w^{y,y}\over 2} \nonumber \\
\sigma_L & = & {4\pi^2\alpha\over K}{{\bf q}^2\over Q^2}
\left[w^{0,0}-2{q_0\over {\bf q}}w^{0,q} + 
({q_0\over {\bf q}})^2w^{q,q}\right] 
\nonumber \\
\sigma_{TT} & = & {4\pi^2\alpha\over K}{w^{x,x}-w^{y,y}\over 2} \nonumber \\ 
\sigma_{TL} & = & {4\pi^2\alpha\over K}\sqrt{{2{\bf q}^2\over Q^2}}
\left[{q_0\over {\bf q}}w^{q,y} - w^{0,y}\right] 
\label{sfun}
\end{eqnarray}
Thus, knowledge of the electromagnetic transition 
amplitude $F^{\mu}$ will allow us to calculate all of the above 
structure functions and  the differential cross section of eq.(\ref{crs2}).
To proceed, we  express  $F^{\mu} = F^{\mu}_{a} + F^{\mu}_{b}$. 
Here, $F^{\mu}_{a}$ describes a transition amplitude within the impulse 
approximation when the resonance produced on one nucleon 
does not experience any further interaction (see Fig.1a), and
$F^{\mu}_{b}$ describes the amplitude where 
additional rescattering of the electromagnetically produced 
hadronic system is taking place (see Fig.1b).
Next we shall outline the calculation of the $F^{\mu}_{a}$ and 
$F^{\mu}_{b}$  amplitudes.

\subsubsection{Impulse approximation}

The scattering amplitude  within the impulse approximation (IA) 
corresponds to the diagram of Fig.1a, where the $S_{11}$ produced by the 
electromagnetic interaction does not  interact further with spectator 
nucleon
\begin{equation}
F_a = (2\pi)^{3\over 2}\psi(p_s)A^{\mu}(Q^2),
\label{Fa}
\end{equation}
where $A^{\mu}(Q^2)$ is the electromagnetic $\gamma N N^*$ transition 
amplitude and   $\psi$ is the nonrelativistic deuteron wave function 
normalized as $\int|\psi|^2(p)d^3p=1$.

\subsubsection{Rescattering amplitude}

We next consider the rescattering amplitude of Fig.1b, where the 
hadronic system ($h$) produced in an electromagnetic scattering on one nucleon 
rescatters off the second spectator nucleon producing the $S_{11}$ in the 
final state.  
Suppressing the spin indices, the rescattering amplitude  can be 
represented as follows (see e.g. \cite{FMGSS,FSS97}):
\begin{equation}
F^\mu_{b}=  {1\over \sqrt{2m}}\sum\limits_{h}\int 
{A^{\mu}_h(Q^2) \Gamma(p_d,p'_s) f^{hN\rightarrow N^*n}(p_f,p'_s,p_s)\over 
[(p_d-p'_s)^2-m^2+i\epsilon] [p'^2_s-m^2+i\epsilon]
[(p_f+ p_s - p'_s)^2 - m_{h}^2 + i\epsilon] } 
{d^4p'_s\over i(2\pi)^4},
\label{Fb1}
\end{equation}
where, $p'_s$ and $p_{s}$ are the momenta of the spectator nucleon in the 
intermediate and final state, respectively, $p_f$ is the momentum of the 
$S_{11}$ in the final state, $\Gamma(p_d,p'_s)$ is the 
invariant vertex of the transition $d\rightarrow pn$ into two off-mass shell 
nucleons, and  $f^{hN\rightarrow N^*N}$ are the  $hN\rightarrow S_{11}N$ 
diffractive transition  amplitudes.  All  spin dependences of the target 
nucleons are included in the vertex factor. Here ${1\over \sqrt{2m}}$ 
arises from the normalization of the spectator nucleon wave function.
Using a non-relativistic description of Fermi motion in the deuteron 
allows us to evaluate the loop integral by taking the residue over the 
spectator nucleon energy in  the intermediate state, i.e. we can replace 
$[p'^2_s-m^2+i\epsilon]^{-1}d^0p'_s$ by $-i(2\pi)/2E'_s\approx -i(2\pi)/2m$. 
This is possible because, in this case, there is 
only one nearby pole in the lower part of the $p'_{s0}$ complex plane 
(see for details Refs.(\cite{FMGSS,FSS97})).

The calculation of the residue in  $p'_{s0}$ fixes the time ordering from 
the left to the right in diagram Fig.1b. We introduce the nonrelativistic 
deuteron wave function as 
$\psi(p_d-p'_s)\equiv {\Gamma^{d\rightarrow pn}
\over [(p_d-p'_s)^2-m^2+i\epsilon]
\sqrt{(2\pi)^32m}}$ (with $\int|\psi(k)|^2d^3k=1$).
Performing above integration we obtain
\begin{eqnarray}
F^\mu_{b} & = &  -{(2\pi)^{{3\over 2}}\over 2m}
\sum\limits_h\int A^{\mu}_{h}(Q^2)\psi(p'_s)  
{f^{hN\rightarrow N^*N}(p_f,p'_s,p_s)\over 
[(p_f+ p_s - p'_s)^2 - m_{h}^2 + i\epsilon] } {d^3p'_s\over (2\pi)^3}  
\nonumber \\ & = &
-{(2\pi)^{{3\over 2}}\over 2} \sum\limits_{h}\int A^{\mu}_{h}(Q^2)\psi(p'_s) 
{f^{hN\rightarrow N^*N}(p_{st}-p'_{st})\over 2mp_{fz} [ p'_{sz}- p_{sz} + 
\Delta  + i\epsilon] } {d^3p'_s\over (2\pi)^3},
\label{Fb2}
\end{eqnarray}
where 
\begin{equation} 
\Delta = (E_s-m){E_f\over p_{fz}} - (p_{st}-p'_{st}){p_{ft}\over p_{fz}} 
+ {W^2 - m_h^2\over 2 p_{fz}},
\label{delta}
\end{equation}
where $W=p^2_{f}$ defined according to eq.(\ref{W2}).
In the last part of eq.(\ref{Fb2}) we used energy-momentum conservation  
to express  the propagator of the hadrons $h$ produced in the intermediate  
state as:
\begin{eqnarray}
& & (p_f+ p_s - p'_s)^2 - m_{h}^2 =   \nonumber \\
& &  2p_{fz}\left[p'_{sz}- p_{sz}  + (E_s-m){E_f\over p_{fz}} - 
(p_{st}-p'_{st}){p_{ft}\over p_{fz} } + 
{(p_s - p'_s)^2\over 2p_{fz}} +{W^2 - m_{h}^2\over 2p_{fz}} \right]  
\nonumber \\
& & \approx  2p_{fz}[p'_{sz}- p_{sz} + \Delta].
\label{kin}
\end{eqnarray}
The fact that the  energy  transferred in the soft 
$hN\rightarrow N^*N$ rescattering is small compared to the total 
energy of the scattered particles allows us to neglect the term  
${(p_s - p'_s)^2\over 2p_{fz}}$ ($\sim 
{(\vec p_s - \vec p_s\/')^4\over 8 m^2p_{fz}} - 
{(\vec p_{s}-\vec p_{s}\/')^2\over 2p_{fz}}$ with  
$|\vec p_{s}-\vec p_{s}\/'|\approx 0.2~GeV/c$)
as compared to the other contributions  to $\Delta$. 

We keep the term $(E_s-m){E_f\over p_{fz}}$ because it does not vanish 
with an increase of the projectile energy at fixed  
spectator nucleon momentum. The term $(p_{st}-p'_{st}){p_{ft}\over p_{fz}}$ 
vanishes for the kinematics being considered, where $p_f$ is nearly parallel 
to the momentum of virtual photon. The last term proportional to 
${m_{N^*}^2 - m_{h}^2\over 2p_{fz}}$ takes into account minimal longitudinal 
momentum need to be transfered to make a nondiagonal $h-N^*$ transitions. 
Because this term enters as an effective longitudinal momentum into 
the argument of the deuteron wave function it suppress the contributions of 
intermediate states $h$ with masses far from $m_{N^*}$.
   
The fact that the soft scattering amplitude depends only weakly  on 
the initial energy helps to simplify  eq.(\ref{Fb2}). It is convenient to  
redefine the soft scattering amplitude 
${f^{hN\rightarrow N^*n}(p_f,p'_s,p_s)\over 2p_{fz}m} 
\approx  f^{hN\rightarrow N^*n}(p_{s}-p'_{s})$,
where  $f^{hN\rightarrow N^*n}$
is now the soft scattering  amplitude normalized similarly to  the elastic
amplitude, which is in turn normalized by the optical theorem
$Im f^{hN\rightarrow hN}(k=0)=\sigma_{hN\rightarrow hN}^{tot}$.

Then,  introducing the transfered momenta 
$k=p_s-p'_s$,  eq.(\ref{Fb2}) can be rewritten as  
\begin{equation} 
F_{b}^{\mu}   =    - {(2\pi)^{{3\over 2}}\over 2}
\sum\limits_{h}\int \psi_d(p_s-k)  A^{\mu}_h(Q^2)
{ f^{hN\rightarrow N^*N}(k)\over 
[-k_z + \Delta  + i\epsilon] } {d^3k\over (2\pi)^3}.
\label{Fb3}
\end{equation}
In eq.(\ref{Fb3}), the only quantities to be specified  
are the electromagnetic transition amplitude  $A^{\mu}_h(Q^2)$ and 
the soft rescattering amplitude $f^{hN\rightarrow N^*N}(p_{s}-p'_{s})$.

In next two sub-sections we will discuss the constituent quark model 
and the chiral dynamic approach and perform the calculation of the impulse 
approximation ($F^{\mu}_a$) and rescattering ( $F^{\mu}_b$) amplitudes
in each of them.

\subsection{Predictions within CQM approach}

Within the constituent quark model approach, we assume that the 
intermediate states are either nucleons ($N$) or an $N^*$ resonance, 
whose structure is described by the CQM.  Furthermore, we neglect the 
$NN\rightarrow N^*N$ transition amplitude compared to the amplitude of 
elastic $N^*N\rightarrow N^*N$ scattering. The relative suppression of the 
transition amplitude compared to the elastic amplitude is supported by 
experimental observation
in Ref.~\cite{transition}, which gives upper limit of such suppression as 
${\sigma_{NN\rightarrow N^*N}\over \sigma_{N^*N\rightarrow N^*N}} 
\sim {1\over 20}$.  Another source of suppression is our choice of 
kinematics in eq.(\ref{xbj}). Because this  corresponds to the  
production of an $N^*$ resonance off a quasifree  nucleon
and $x<1$, the struck nucleon ($N$) in the initial state is highly virtual.
 The resulting phase change in the amplitude is
$\exp(i\Delta E z)$, where according to eq.(\ref{delta}) $\Delta E\approx 
{m_{N^*}^2-m^2\over 2 p_{fz}}$ accounts for the virtuality of the intermediate 
nucleon.  Since this phase will contribute a longitudinal component of
momentum into the deuteron wave function, one can estimate the  suppression as 
${|\psi_{D}(\Delta E)|\over|\psi_{D}(p\approx 0)|}$.

One gains a further simplification by neglecting the charge exchange 
contribution in the soft rescattering amplitude.  Neglecting this is justified 
by the fact that the charge-exchange amplitude is predominantly real due to 
its pion-exchange nature. Thus, it will interfere mainly with the real part of
the diffractive $N^*N$ amplitude, which is $\le 5\%$ of the total cross 
section (see discussion in Ref.\cite{FMGSS}).

The two approximations mentioned above will allow us to factorize the 
electromagnetic amplitude in $F^{\mu}_b$ so that we finally obtain:
\begin{equation}
F^{\mu}_b=    -A^{\mu}_{N^*}(Q^2){(2\pi)^{{3\over 2}}\over 2}\int \psi(p_s-k)  
{ f^{N^*N\rightarrow N^*N}(k)\over 
[ -k_z + \Delta  + i\epsilon] } {d^3k\over (2\pi)^3}.
\label{Fbcqm}
\end{equation}
Here
\begin{equation} 
\Delta = (E_s-m){E_f\over p_{fz}} - (p_{st}-p'_{st}){p_{ft}\over p_{fz}} 
+ {W^2 - m_{N^*}^2\over 2 p_{fz}} + i\Gamma m_R/(2p_{fz}),
\label{deltacqm}
\end{equation}
where the term $i\Gamma m_R/(2p_{fz})$ accounts for the mass width of the
resonance produced in the intermediate state (see e.g. \cite{RF}).
The size of this correction decreases with increasing energy, and in the
high-energy limit the location of the pole will be again defined by the mass
of the propagating resonance.

In eq.(\ref{Fbcqm}) $f^{N^*N\rightarrow N^*N}$ is the small angle elastic 
scattering amplitude, which can be represented in the 
form (see e.g. Ref.\cite{AG}):
\begin{equation}
f_{N^*N\rightarrow N^*N}(t) \approx  
\sigma^{tot}_{N^*N}(i + \alpha) e^{{b\over 2}t},
\label{ampl}
\end{equation}
where $\sigma^{tot}_{N^*N}$ is the total $N^*N$ scattering cross section, 
$b$ defines the slope factor of the elastic differential cross section,
and $\alpha$ accounts for the real part of the amplitude.

To construct the soft amplitude $f^{N^*N\rightarrow N^*N}$ within the 
CQM framework, the idea is to exploit the fact that in general the small-angle 
elastic $hN$ scattering depends on the radius of the hadron in a 
characteristic fashion.  In particular, the hadronic radii of the interacting 
particles are related (see e.g. Refs.\cite{GS,PH,LW}) to the total scattering 
cross section $\sigma^{tot}_{hN}$ as follows,
\begin{equation}
\sigma^{tot}_{hN} = \sigma^{tot}_{NN}
{<r^2_{h}>\over <r^2_{N}>},
\label{stot}
\end{equation}
and to the slope parameter $b$,
\begin{equation}
b \approx {1\over 3} \left(<r^2_{h}> +  <r^2_{N}>\right),
\label{slope}
\end{equation}
appearing in Eq.(\ref{ampl}).  It follows from eqs.(\ref{ampl}), (\ref{stot}) 
and (\ref{slope}) that if we implement elastic $S_{11}N$ scattering 
based on the well established characteristics of the nucleon 
we can establish the spatial parameters of the $S_{11}$ resonance.

One may now recall that within the CQM, the mean-square radius of a quark 
orbit, $<r^2>$, scales roughly as $2N+L+3/2$, where $N$ is the number of 
radial nodes and $L$ is the orbital angular momentum in the wave function.
Thus, the spatial distribution of the quarks is quite sensitive to their 
orbital excitation within the individual resonances.  

To estimate the dependence of the reaction on the size of the hadrons in
the CQM, it is first necessary to eliminate the dependence on
the center-of-mass coordinate in the hadron wave function.  In the 
zero-order quark shell model of  Ref.~\cite{size1} this wave function is 
$\Psi = \phi(N_\rho,L_\rho,\rho)  \phi(N_\lambda,L_\lambda,\lambda)$, where 
$\rho^2 = (r_1-r_2)^2/2$ and $\lambda^2 = (r_1+r_2-2r_3)^2/6$ 
with $r_1$, $r_2$, and $r_3$ being the coordinates of the 
three constituent quarks.  The quantities $N_\rho$ and $L_\rho$ represent 
the radial and orbital excitation quantum
numbers. One then obtains the following relation for the radius of the baryons
in terms of the mean-square radii of the two independent harmonic oscillators:
\begin{equation}
<r^2>=[(2N_\lambda+L_\lambda+3/2)+(2N_\rho+L_\rho+3/2)]b_{hosc}^2.
\label{CQMR}
\end{equation}
where $b_{hosc}^2$ is the slope factor of the Harmonic Oscillator wave function
of constituent quark. For the nucleon, $(N,L)=(0,0)$ for both sets of quantum numbers, and 
$<r_{nucl}^2> = 3b_{hosc}^2$. For the $S_{11}$ one set of $(N,L)=(0,0)$ 
and the other is $(0,1)$.  This gives $<r_{s11}^2> = 4b_{hosc}^2$.
Therefore according to eqs.(\ref{stot},\ref{slope}) one obtains:
\begin{equation}
(\sigma^{tot}_{hN},b)= ({4\over 3}\sigma^{tot}_{NN},{7\over 6} b_{NN}).
\label{CQMP}
\end{equation}
The coefficients on the right-hand side may be larger than those
given here, since the harmonic oscillator model may overestimate
the effect of confinement.  For a sufficiently sensitive experiment
of the type we propose here, one should be able to determine the extent to
which the $S_{11}$ resonance is larger than a nucleon.

\subsection{Predictions  within the ECL approach}

Within the chiral SU(3) dynamics approach, the $S_{11}$ represents a 
superposition of $N\pi$, $N\eta$, $\Lambda K$, and $\Sigma K$ states
with total isospin${1\over 2}$\cite{KSW}. One can therefore describe the 
intermediate state of Fig. 1b  by these four states, which then 
interact with the spectator nucleon. Such a picture will correspond to the 
following rescattering amplitude in Eq.(\ref{Fb3}):
\begin{equation} 
F_{b}^{\mu}   =    - {(2\pi)^{{3\over 2}}\over 2}
\sum\limits_{i}\int \psi(p_s-k)  A^{\mu}_i(Q^2)
{ f^{iN\rightarrow jN}(k)\over 
[ -k_z + \Delta_{ij}  + i\epsilon] } {d^3k\over (2\pi)^3},
\label{Fbecl}
\end{equation}
where $i,j = (1-4)$ represent the four relevant meson-baryon channels 
states: $N\pi$, $N\eta$, $\Lambda K$, and $\Sigma K$, respectively, 
and $A_i^{\mu}$ are the amplitudes for the corresponding electromagnetic 
transitions  $\gamma N\rightarrow N\pi, N\eta, \Lambda K, \Sigma K$.
We take $\sum\limits_i \equiv \sum\limits_i \int {d^3k_i\over (2\pi)^3}$, 
with the $1/2\sqrt{m_i^2+p_f^2}$ term absent because of the normalization 
used for the rescattering amplitude (see before of the eq.(\ref{Fb3})). 
The quantity $\Delta_{ij}$, which accounts for the longitudinal momentum 
transfer is:
\begin{equation} 
\Delta_{ij} = (E_s-m){E_f\over p_{fz}} - (p_{st}-p'_{st}){p_{ft}\over p_{fz}} 
+ {m_{i}^2 - m_j^2\over 2 p_{fz}},
\label{deltaij}
\end{equation}
where $m_i = (\sqrt{m_M(i)+k_i}+ \sqrt{m_B(i)+k_i})^2$ is the off-shell 
invariant mass of the intermediate system. Accordingly, $m_j\equiv W$ is 
the mass of the final state hadronic system.

In eq.(\ref{Fbecl}), $f^{iN\rightarrow~jN}(q)$ corresponds to the scattering 
of the initial correlated meson-baryon pair "$i$"  off the spectator 
nucleon to the final correlated meson-baryon pair in state "$j$" with the 
spectator nucleon recoiling. The amplitude of such a scattering can be 
represented through the transition form-factors $S_{i,j}(q)$ (see e.g. 
Refs.{\cite{FG,FKMPSS}) as follows:
\begin{eqnarray}
f^{iN\rightarrow jN}(k) & = &  f^{MN}(k)S_{i,j}(a_1\cdot k) + 
f^{B N}(k)S_{i,j}(-a_2\cdot k) +   \nonumber \\
& & {i\over 2}\int {d^2k'_{\perp}\over (2\pi)^2} 
f^{MN}(a_1\cdot k - k')f^{B N}(a_2\cdot k + k')S_{i,j}(k'_{\perp})
\label{CDP}
\end{eqnarray}
where $a_1={m_{B}\over m_{B}+m_{M}}$ and $a_2={m_{M}\over m_{B}+m_{M}}$ 
($M$ and $B$ defines the meson and the baryon  which belongs to the
intermediate ``i'' state) $f^{MN}$ and $f^{BN}$ are the diffractive 
amplitudes of the meson-nucleon and baryon-nucleon small angle scattering,  
defined in the form of eq.(\ref{ampl}). The $S_{i,j}(q)$ is the transition  
structure function of the meson-baryon system:
\begin{equation}
S_{i,j}(k) = \int d^3 r \psi_{k_i}(r)\psi^{\dag}_{k_j}(r)e^{-i\vec k \vec r},
\label{SF}
\end{equation}
where $\psi_{ki}(r)$ is meson-baryon wave function for channel $i$, which 
can be expressed (see e.g. \cite{BJ}) as follows
\begin{equation}
\psi_{k_i}(r) = \phi^i_{k_i} +  
\sum\limits_{n}\int {d^3l\over (2\pi)^3} 
{\phi^n_{k_n} <l|T_{ni}|k_i>\over k_n^2 - l^2 + i\epsilon}.
\label{wfij}
\end{equation}
Here, the on-mass shell momentum is defined as
$k_n^2 = {[W^2-(m_B(n)+m_M(n))^2][W^2-(m_B(n)-m_M(n))^2]\over 4W^2}$.
In eq.(\ref{wfij}) $\phi^i$ is the plane wave function for state $i$
and  $T_{ni}$ is the t-matrix of the $n\rightarrow i$ transition, which 
represents the solution of the coupled channel Lippmann-Schwinger equation
\cite{KSW,KWW}. In the calculations, we made the partial wave decomposition 
of the wave function in eq.(\ref{SF}) using the relation 
(for any operator $A$)\cite{LL}:
\begin{equation}
<k'|A|k> =  4 \pi \sum\limits_{l}(2l+1)P_l(k',k)<k'|A_l|k> ,
\label{pwl}
\end{equation}
retaining only the $S$-wave ($l=0$) contribution.   Such a restriction is 
justified by the fact that 
eq.(\ref{CDP}) corresponds to small-angle scattering,
where the main contribution comes from momenta $k\le 200 MeV/C$. 
Since $q$ enters into $S_{ij}$ as $a_{1,2}k$, even smaller
momenta are relevant for the rescattering amplitude.

Finally we note that for  both the electromagnetic amplitude 
$A^{\mu}_i(Q^2)$ and the matrix $T_{ni}$ we use the calculation of 
the Ref.\cite{KWW,TW}.

\section{Numerical estimates}

For the numerical results we present in this section, we assume that the 
quantities measured are the momenta of the final electron and the spectator 
neutron.

To assess to what extent the hadronic structure of $S_{11}$
is revealed in the rescattering processes, we must first ensure that the 
rescattering (amplitude $F^\mu_b$) dominates in the overall scattering 
amplitude. For this purpose we will consider the kinematics 
close to the condition described in Sec. II.A. Within this 
kinematics one considers the  following ratio:
\begin{equation}
R = {\sigma(Q^2,W,\vec p_{s})\over \sigma^{0}(Q^2,W,\vec p_{s})},
\label{R}
\end{equation}
where $\sigma$ is the differential cross section of eq.(\ref{crs1}) 
that includes both impulse approximation and the rescattering amplitudes,
$F^\mu_a$ and $F^\mu_b$, respectively.  The cross section $\sigma^0$ 
corresponds to the impulse approximation only. Because of the destructive 
character of the interference between the impulse approximation and 
rescattering amplitudes above, the ratio has a functional form 
$R\sim 1 - 2{|F_aF_b|\over |F_a|^2} +{|F_b|^2\over |F_a|^2}$. 
As follows from eq.(\ref{Fb3}), the deuteron wave function in 
$F^\mu_b$ enters as $\psi_d(p_{s}-q)$, compared to $\psi_d(p_s)$ 
in the impulse approximation of eq.(\ref{Fa}). Because of the different 
arguments appearing in the deuteron wave function, by increasing $p_{s}$ 
one should in general expect a more dominant contribution from $F^\mu_b$ 
since it contains the loop integration with effective momenta 
$|\vec p_s - \vec q|\le p_s$.

The analysis of the Ref.\cite{FMGSS} demonstrates that the similar ratio 
for quasielastic $d(e,e'N)N$ scattering exhibits a strong dependence on 
the spectator nucleon momentum $\vec p_s$. With the increase of $p_s$, 
$R$ first  decreases below one (because of the dominant contribution from 
interference term $|F_aF_b|$, usually called the {\em screening} effect). 
Then it increases above one as the {\em double scattering} contribution
$|F_b|^2$ becomes dominant.
 
In Fig.~2 we demonstrate the dependence of $R$ on $\theta_{sq}$, the polar
angle of the spectator nucleon with respect to the momentum of the virtual 
photon. We fix $Q^2=1~GeV^2$, $W=1.54~GeV$ and chooses two characteristic 
values for the momenta $p_s$ ( $200~MeV/c$ and $400~Mev/c$), where,  
respectively, the screening and double scattering terms are important. 
Note that the minimum in Fig.~2a  and the maximum in Fig.~2b correspond to the 
value of $x$ and ${E_s-p_{sz}\over m}$ defined by the conditions of 
eq.(\ref{xbj}) and eq.(\ref{alpha}), which ensures the maximal contribution 
from rescattering amplitude $F^\mu_{b}$ (Fig.~1b).

The solid line in Figure 2 corresponds to the calculation within the CQM,
where we assume that the hadronic size of the $S_{11}$ is the same 
as for the nucleon. The dashed line also corresponds to the calculation within 
the CQM model, but the radius of the $S_{11}$ is described using the relation 
of eq.(\ref{CQMR}) with the parameters of the rescattering  amplitude 
defined according to eq.(\ref{CQMP}).  Note that whereas our CQM calculation 
predicts $15\%$ more screening for the larger $S_{11}$ at $p_s=200~Mev/c$, 
it reveals more than $50\%$ greater rescattering in the kinematics where 
double scattering dominates, $p_s= 400~MeV/c$.  Thus, such 
an increase of $R$ with the spectator momentum could clearly indicate a
large radius of the resonance.

A qualitatively similar picture is obtained for the calculation of 
$R$ within ECL approach, only now the angular distribution is 
somewhat broader because of the nondiagonal hadronic state contributions.
This is because intermediate states with different mass contribute with 
different longitudinal momentum transfer due to the term  
$\sim {m_{i}^2-m_j^2\over 2 p_{fz}}$ in eq.(\ref{deltaij}) .

A general feature of the $R$ is that at $p_{st}< 300~MeV/c$ 
the screening effects are dominating in eq.(\ref{R}), thus $R<1$, 
and at $p_{st}\ge300~MeV/c$ the dominant character of the double 
scatterings makes $R>1$.  Such a trend suggests that one can 
introduce another ratio $R_{\sigma}$, which corresponds 
to the ratio of the cross section measured say at $p_{s}\ge 400~MeV/c$ 
to the cross section  measured at $p_s\approx 200MeV/c$:
\begin{equation}
R_{\sigma}(p_{s1},p_{s2}) = 
{\sigma(p_s\approx 400 MeV/c)\over \sigma(p_s\approx 200 MeV/c)}.
\label{R12}
\end{equation}
Because of the different trends of the prediction at the two kinematic 
ranges, this ratio becomes more sensitive to the hadronic 
structure of the reinteraction than each cross section does separately.

In Figure 3 we represent the angular dependence of $R_{\sigma}(p_{s1},p_{s2})$
normalized by $R_{\sigma_0}$ calculated for $p_{s1}=400$ and 
$p_{s2}=200~MeV/c$. As follows from this figure, the CQM predictions 
corresponding to the larger resonance radius of eqs.(\ref{CQMR}) and
(\ref{CQMP}) differ by a factor of 2 from those corresponding to an $S_{11}$ 
whose radius has been taken equal to the nucleon radius.

Next, we will consider another measurable characteristic which could 
be complementary to that given above. This is the $W$ dependence (mass 
distribution) of the cross section at fixed ${E_s-p_{sz}\over m}=1$ 
(eq.(\ref{alpha})) and different values of $p_{st}$.  

Within the ECL approach, where the $S_{11}$ represents the superposition of
four meson-baryon isospin-${1\over 2}$ states, one expects a larger 
contribution of the higher-mass intermediate states in the rescattering 
amplitudes with an increase of $W$. Specifically, as follows from 
eq.(\ref{deltaij}), with an increase of $W$, the contribution of the more 
massive $K\Lambda$ and $K\Sigma$ intermediate states will be the least 
suppressed by the longitudinal momentum.  As a result, one may expect that 
the final state interaction will grow with increase of $W$. 

In Fig.~4 we present the $W$ dependence of the  $\sigma_T(d(e,e'\eta p)n)$
cross section calculated according to the eq.(\ref{sfun}) within the ECL 
approach.  The cross sections are  normalized by the square of the deuteron 
wave function $|\psi_d(p_s)|^2$ and by values of $R$ calculated at 
$W=1.54~GeV$. The calculations are done at $Q^2=1~GeV^2$ for fixed 
${E_s-p_{sz}\over m}=1$ and for different values of $p_{st}$. The figure shows 
little deformation of the mass distribution for kinematics where the 
rescattering results from the screening effect (at $p_{st}\le 200 ~MeV/c$).  
This reflects the fact that while the real part of the rescattering amplitude 
makes the $W$ distribution broader, the increased  contribution of the 
intermediate masses into the imaginary part of the rescattering amplitude 
results in a  sharpening of the $W$ distribution (because of more absorption 
at $p_{st}\le 200 ~MeV/c$). Thus these two effects tend to cancel each 
other, and the resulting distribution is less affected by the final state 
interaction.

For kinematics where double scattering dominates, both the real and the
imaginary parts of the rescattering amplitude work in the same direction,
to broaden the $W$ distribution. Here one observes a  substantial 
broadening of the mass distribution. Such a broadening is the essential 
signature of the composite nature of the resonance.
  
Figure 5 presents the analogous $W$ dependence of $\sigma_{tot}(d(e,e'\eta p)n)
= 4\pi(\sigma_{T} + \epsilon \sigma_{l})$ calculated within the CQM approach.
For the calculation of the electromagnetic transition part of the cross 
section $\gamma^*N\rightarrow N\eta$, we used the parameterization of 
$\sigma_{tot}(\gamma^*p\rightarrow p\eta)$ from Refs.\cite{Br}.  As in the case 
of Figure 4, the calculations were done at $Q^2=1~GeV^2$ for fixed 
${E_s-p_{sz}\over m}=1$ and different values of $p_{st}$.

As follows from eq.(\ref{Fbcqm}) and eq.(\ref{deltacqm}), the $W$ dependence
of the FSI amplitude is mainly due to the term 
${W^2 - m_{N^*}^2\over 2 p_{fz}}$.  Because the kinematics was chosen with 
${E_s-p_{sz}\over m}=1$, such dependence will suppress the FSI amplitude at 
$W>m_{N^*}$ ($W~<~m_{N^*}$). Thus one would  have less FSI in the tails of 
the $W$ distribution.  Note that because $p_{fz}$ grows with $W$, some 
additional
$W$ dependence comes from the term  $i\Gamma m_R/(2p_{fz})$, which will
slightly shift the maximum of the FSI to larger $W$.

Because of the suppressed FSI in the tails of $W$ distribution, one observes a
broadening of the overall $W$ distribution for kinematics where the FSI 
results from screening ($p_{st}\le 0.2~Gev/C$) compared to the $W$ 
distribution within the IA.  However, for the kinematics of double scattering,
$p_{st}\ge 0.3 Gev/c$ where the FSI becomes dominant, the $W$ dependence of 
the FSI has the opposite effect on the overall $W$ dependence of the 
cross section compared to the IA contribution.

Comparing Figs. 4 and 5, one concludes that the $W$ dependence of the FSI is 
opposite within the ECL and CQM.  Within the ECL, at larger $W$ one has a larger
FSI because of the increased contribution of the large mass intermediate 
states. On the other hand, within the CQM, the FSI is smaller at larger $W$ 
because of the larger longitudinal momentum transfer entering into the FSI 
amplitude.  Note that this sensitivity of the FSI on $W$ will be 
suppressed with an increase of $Q^2$ ($p_{fz}$).  The suppression will be more 
pronounced within the CQM, since the $W$ dependence is determined mainly by  
the factor $\Delta$ of eq.(\ref{deltacqm}).

\section{Further problems that can be addressed}

\noindent{\em Class of the resonances which can be investigated by similar  
 reactions:}

\medskip

{}From the discussion of the previous section, we see that the radius of a
baryon resonance, which scales sensitively according to the orbital quantum 
numbers of the quarks in it in a definite model (e.g., that of 
Ref.\cite{size1}), bears a rather straightforward relationship to the 
quantum numbers of the resonance itself.  Since the electro-production of a
baryon resonance in the type of experiment we study is sensitive to its 
mean-square radius, and since there is a rich variety of such baryon 
resonances with various quantum numbers, the opportunity arises to 
determine experimentally whether a scaling relationship based on these 
simple ideas describes the structure of the baryon resonances.  

Whereas we have discussed the size determination only in the context of the 
$S_{11}$, extending the study to other resonances is straightforward within
the same theoretical framework.  The extent to which an experimental study
would be feasible for other resonances will depend on the rates, which in turn 
will depend upon the magnitude of the coupling of the photons to the baryon 
resonances in question, and the size of the theoretical background amplitudes 
for $NN\rightarrow NN^*$ and charge-exchange amplitudes. An examination of
the data on resonance production~\cite{transition} shows that the same upper
limit applies for all other baryon resonances studied as it does to the
$S_{11}(1535)$.  The nondiagonal transition cross section will be even more 
suppressed for the higher-mass baryon resonances than it is for the 
$S_{11}(1435)$ because of the larger value of $\Delta E$ discussed in 
Sect.~II.C.

By varying the kinematics (e.g., $x$ according to eq.(\ref{xbj})), we have 
the possibility of kinematically selecting the baryon resonance whose size
we are interested in studying.  However, for selecting a given resonance, the 
kinematics 
is not the only relevant consideration. Because there are generally other 
resonances in the same region of $Q^2$ and $q_0$, additional experimental 
constraints are desirable.  For example, in the case of the $S_{11}(1535)$, 
the $D_{13}(1520)$ lies nearby in energy.  The  $S_{11}$ may be 
separated from this by taking advantage of the fact that the $S_{11}$ 
has a relatively large branching ratio for $\eta$ meson decay.  Additionally,
consideration of the electromagnetic form factor shows that requiring 
$Q^2 \ge 1$GeV$^2$ tends to emphasize the  $S_{11}$ over the 
$D_{13}$(1520) \cite{Stoler}.

Because the $D_{13}$ is favored for the smaller $Q^2$ values, this resonance 
may thus be selected by changing $Q^2$.  Like the $S_{11}$, the $D_{13}$ is 
a negative parity resonance and corresponds to a quark excited up by one shell.
In the simple quark model, we would therefore expect the spatial 
distribution of quarks to be similar to that of the $S_{11}$, and hence the
behavior of the reaction cross section $d(e,e'D_{13})n$ should be similar
to that of the $S_{11}$.

The $\Delta_{33}$ resonance would be another interesting case, since in the
simple quark model  all three quarks are in s-wave 
orbits.  Thus, in contrast to the $S_{11}$, the spatial distribution of the 
quarks in the $\Delta_{33}$ is expected to be more compact.  Its reaction 
$d(e,e'\Delta_{33})n$ should therefore show less pronounced final
state interactions than the $S_{11}$, behaving more similarly to the nucleon.
The $\Delta_{33}$ should be easy to detect and analyze since it is a strongly
excited, isolated resonance.  

Another case of interest would be the positive parity $F_{15}(1680)$.
In the simple quark model, this resonance involves the promotion of
one quark two shells up\cite{size1}, and thus the increase in the 
quark spatial distribution would be even more dramatic than in the 
case of the $S_{11}$ or $D_{13}$.  We find for this resonance that 
$(\sigma^{tot}_{hN},b)=(5/3\sigma^{tot}_{NN}, 4/3 b_{NN})$.  
Thus, the interference of the rescattering diagram with the impulse 
approximation  amplitude 
may have a very pronounced signature.  Because of the existence of other 
resonances in the same energy region (such as the $D_{15}$(1675)), 
the $F_{15}$ may be more difficult to separate from the background, 
requiring additional information, perhaps with polarized measurements.  
However, the $F_{15}$ resonance is known to be strongly excited with 
increasing Q$^2$ out to at least 3 GeV$^2$\cite{Stoler}, so it is an 
attractive candidate for study using these same theoretical techniques.
Finally, it is worth mentioning that in case of the production of states
with $L>0$ and $S>1/2$, it would be interesting to try to observe spin effects 
in the rescatterings. In this case, the resonances produced should be
polarized due to the FSI, because of the different color separations for
states (projections) with different helicities (cf. \cite{GFSSG}).

\bigskip

\noindent{\em Relations to other nuclear effects:}

\medskip

It is worth noting that the size of resonances as suggested
here is an important consideration for the interpretation of 
other reactions such as electroproduction\cite{Elouadrhiri} and pion 
scattering\cite{Takahashi} from nuclei in the GeV range of energies,
where resonance excitation is important.  

To unambiguously interpret such experiments, one would like to know, say 
from geometrical considerations, whether the quark distributions of resonances
produced in the interior of nuclei significantly overlap those of the nucleons
of the nucleus.  This has a bearing on whether the resonances can 
be treated as quasiparticle excitations.  The $\Delta_{33}$ is an example of
a resonance believed to
propagate as a quasiparticle, and its medium-modified mass and widths have 
been determined in this situation from meson factory data\cite{Kisslinger}.  
The situation is less clear for higher-mass resonances.  However, theoretical 
analysis of higher energy pion scattering data with such a quasiparticle 
assumption shows that significant medium effects occur for the more massive 
resonances as well\cite{Chen}.

With respect to the quasiparticle assumption, an important  property that 
can be studied in the reactions considered in this paper is the compositeness
of the resonance, or in other words the distribution of the constituents 
that make up the resonant amplitude.  Here one expects a sensitively of the 
mass distribution of final hadronic states to the contribution of 
different intermediate states in the scattering processes. 
Such a sensitivity is substantial at the kinematics dominated 
by hadronic reinteractions. 
Note that the qualitative picture one obtains for the deuteron target 
in rescattering kinematics is similar to the picture of the 
composite hadronic system in nuclear matter\cite{FKW,WL}.

\section{Summary}

The study of the hadronic properties of baryonic  resonances has been carried 
out theoretically using $d(e,e'R)N$ reactions, where the spectator nucleon $N$ 
is detected in a special kinematics that allows substantial 
reinteraction of the electromagnetically produced hadronic 
system with the spectator nucleon.
In particular, the production of the $S_{11}$ resonance 
in the $d(e,e'S_{11})N$ reaction has been studied.

These reactions were shown to be very sensitive to the hadronic 
radius of the resonance.  Measurements such as the ratio of the cross 
sections measured at different values of spectator momenta can yield 
as much as factor 2 difference for the CQM models calculated using 
different assumptions for the spatial distribution of the quarks in the
resonance. In one case we assumed that the  size of the resonance 
was equal to the nucleon size and in another that it  scales according to 
relations suggested by the harmonic oscillator wave function of 
its constituent quarks.

Next we studied the sensitivity of the $d(e,e'S_{11})N$ reactions to the 
composite nature  of the produced resonance. We applied here the ECL 
approach to describe the $S_{11}$ as a superposition of multi-channel 
meson-baryon wave functions with total isospin $1\over 2$. As compared 
to the CQM approximation, the ECL approach predicts a qualitatively 
different picture for the interaction, in that the rescatterings now 
depend on the relative contribution of the different channels in the 
intermediate state.  In particular, one consequence of the composite nature of 
the resonance within the ECL approach is a different pattern (in fact, the 
opposite pattern) of broadening for the mass distribution due to final state 
interactions.

To summarize, the results of the analysis given here suggests that the 
electroproduction of baryon resonances on the deuteron can provide a 
sensitive measure of the hadronic properties of resonances.
This can be achieved by using special kinematics where 
the dominant contribution of the reaction comes from the hadronic 
reinteraction amplitude.
Note that such an experiment might be carried out at TJNAF for 
the case of the $S_{11}$ resonance, for which the theory
and numerical results are worked out in this paper.

\acknowledgments
Very special thanks are due T.~Waas who helped with understanding 
of the details of the ECL calculations and provided the 
computation  of $t$-matrices and electromagnetic amplitudes 
within ECL.
Authors also thankful to N.~Isgur, V.~Burkert, N.~Kaiser 
and G.~Piller for useful discussions. M.Sargsian is grateful 
to  the Alexander von Humboldt foundation for support. 
This works was supported in part by the German-Israeli 
Foundation Grant GIF-I-299.095, the U.S. Department of 
Energy under Contract No. DE-FG02-93ER40771, and the BMBF.

\begin{figure}
\caption{Diagrams corresponding to the reaction 
$e+d\rightarrow e' + N^*+ N$. Kinematics have been chosen in a way 
that suppress the contribution from the $N^*$ component of the deuteron 
ground state wave function.}
\end{figure}

\begin{figure}
\caption{Dependence of $R$ on the angle of the spectator nucleon with respect 
to transfered momentum $q$. Solid lines - calculations where rescattering 
amplitude for $S_{11}$ set as  $NN$ amplitude of elastic scattering. 
Dashed line - $S_{11}N\rightarrow S_{11}N$ amplitude calculated within 
CQM. Dash-dotted curve corresponds to the calculations within ECL 
approach, for $\eta N$ final state. Curves in (a) correspond to the 
spectator momenta $p_s=200~MeV/c$ and in (b) $p_s=400~MeV/c$. 
$W~=~1.54~GeV$ and $Q^2~=1~GeV^2$.}
\end{figure}

\begin{figure}
\caption{Dependence of $R_\sigma$ (normalized by $R_\sigma$ 
calculated within IA) on the angle of the spectator nucleon with respect 
to transfered momentum $q$. The kinematics and definition of the curves 
are  the same as in Fig.2.}
\end{figure}

\begin{figure}
\caption{Dependence of transverse cross section $\sigma_T$ of 
$d(e,e'(\eta N))N$ scattering, on the mass of the produced
hadronic state $W$, calculated within ECL approach. Different
curves correspond to different values of spectator nucleon
transverse momenta $p_{st}$ at fixed ${E_s-p_{sz}\over m} = 1$.
Solid curve corresponds to the $p_{st}~=~0$, dashed curve
- $p_{st}~ =~0.2~GeV/c$, dotted curve $p_{st}~=~0.3~GeV/c$
and dash-doted curve $p_{st} = 400~MeV/c$.}

\end{figure}

\begin{figure}
\caption{Dependence of total cross section
$\sigma_{tot}=4\pi(\sigma_{T} + \epsilon \sigma_L$) of 
$d(e,e'(\eta p))n$ scattering, on the mass of the produced 
hadronic state $W$, calculated within CQM approach.
Different curves correspond to different 
values of spectator nucleon transverse momenta $p_{st}$ at 
fixed ${E_s-p_{sz}\over m} = 1$. Solid curve corresponds to
the $p_{st}~=~0$, dashed curve - $p_{st}~ =~0.2~GeV/c$,
dotted curve $p_{st}~=~0.3~GeV/c$ and dash-doted curve
$p_{st} = 400~MeV/c$.}
\end{figure}

\newpage
\begin{figure}
\centerline{\epsfysize=5.0truein\epsffile{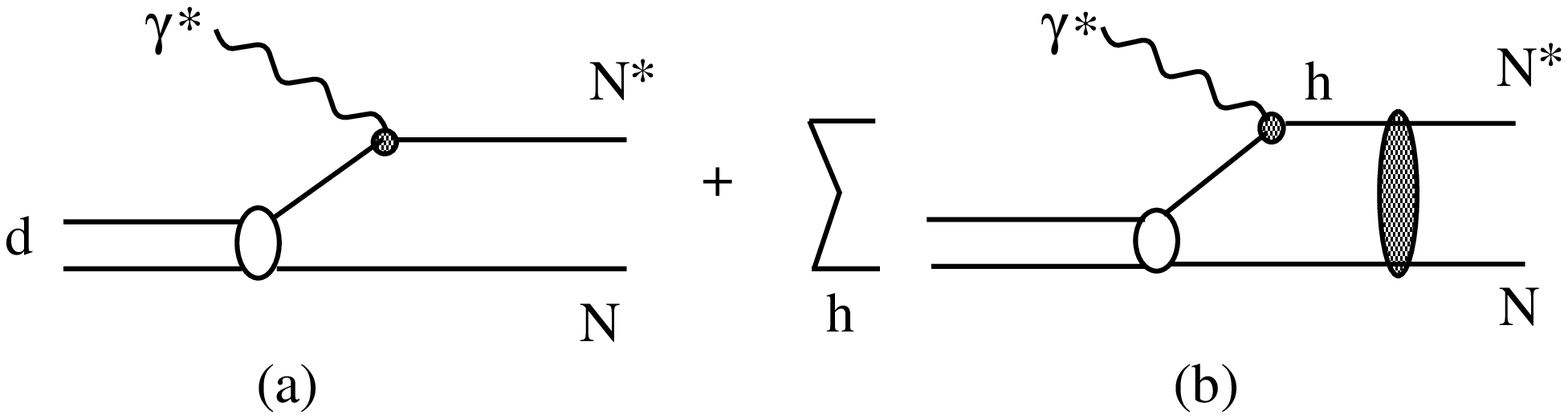}}
\label{Fig.1}
\end{figure}

\noindent{\bf Figure 1}

\newpage
\begin{figure}
\centerline{\epsfysize=7.0truein\epsffile{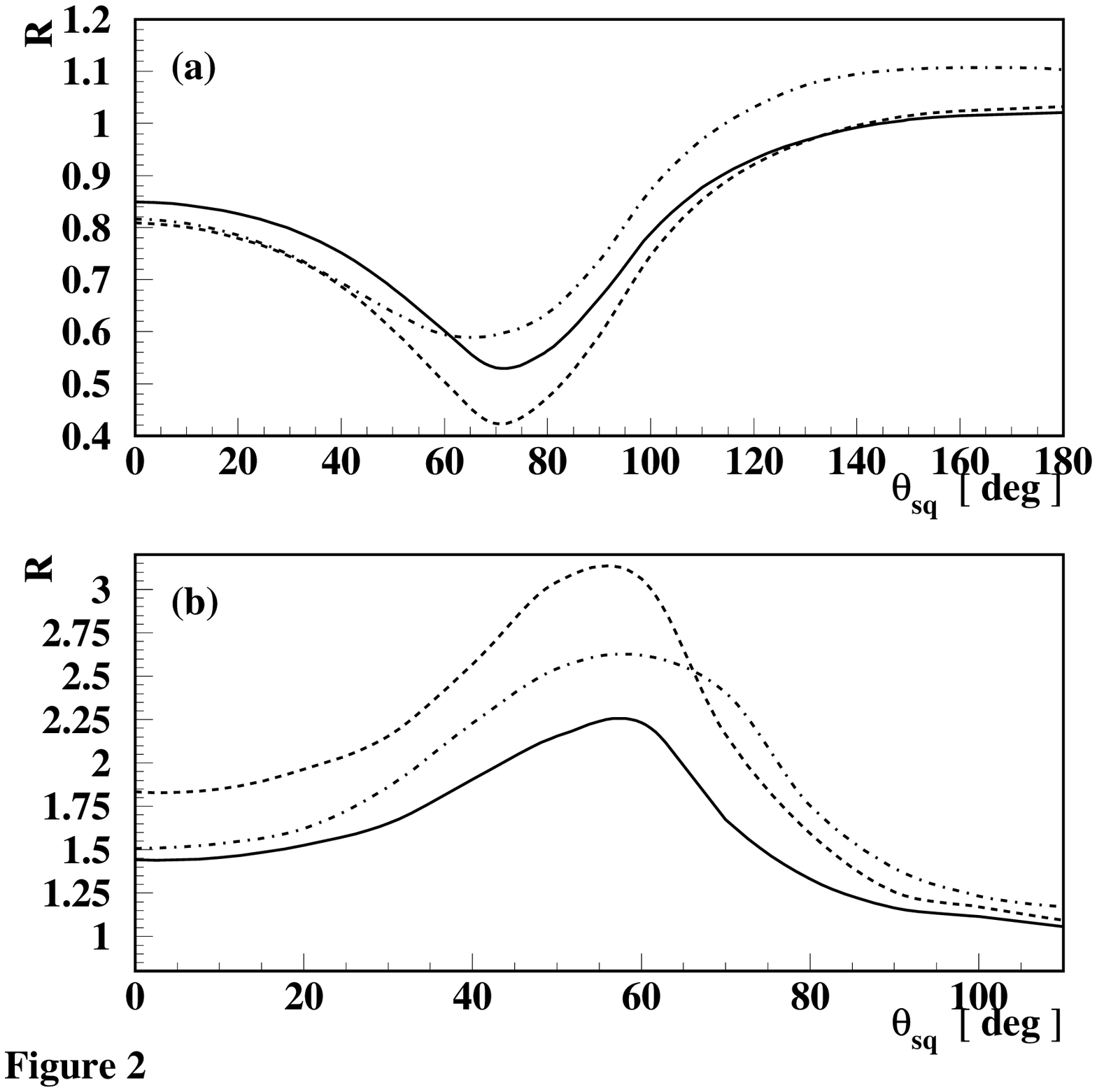}}
\label{Fig.2}
\end{figure}

\newpage
\begin{figure}
\centerline{\epsfysize=7.0truein\epsffile{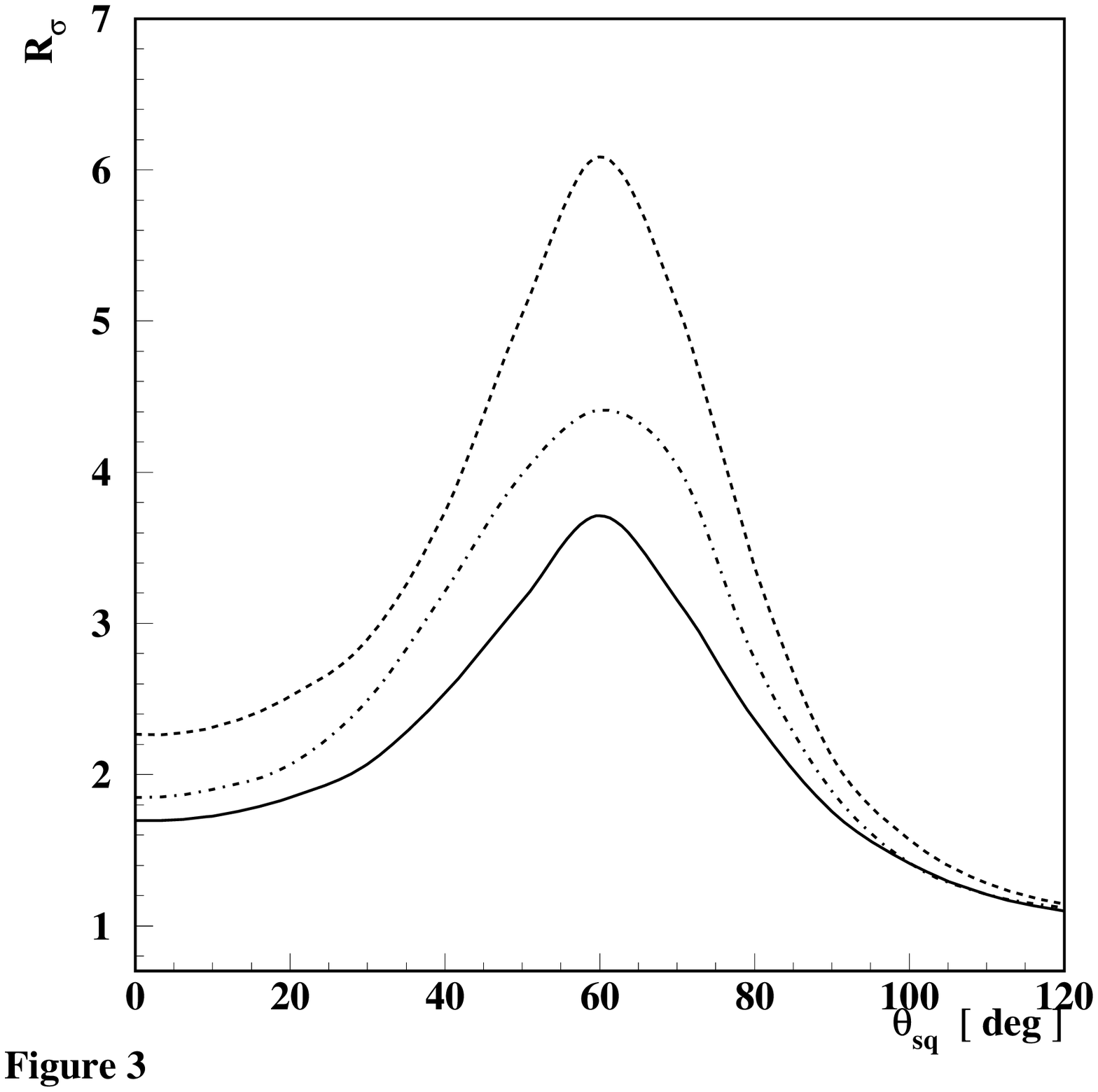}}
\label{Fig.3}
\end{figure}

\newpage
\begin{figure}
\centerline{\epsfysize=7.0truein\epsffile{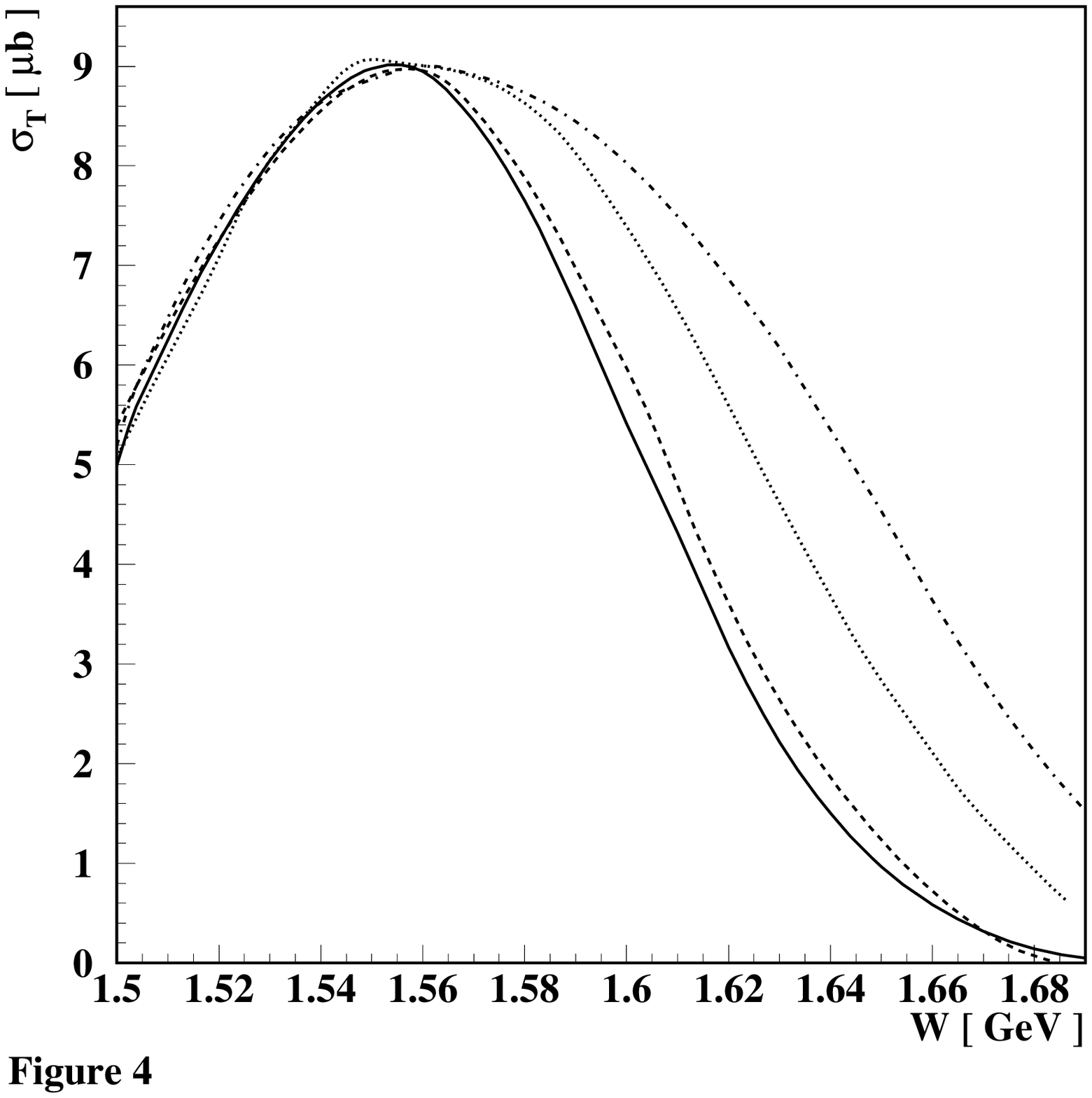}}
\label{Fig.4}
\end{figure}

\newpage
\begin{figure}
\centerline{\epsfysize=7.0truein\epsffile{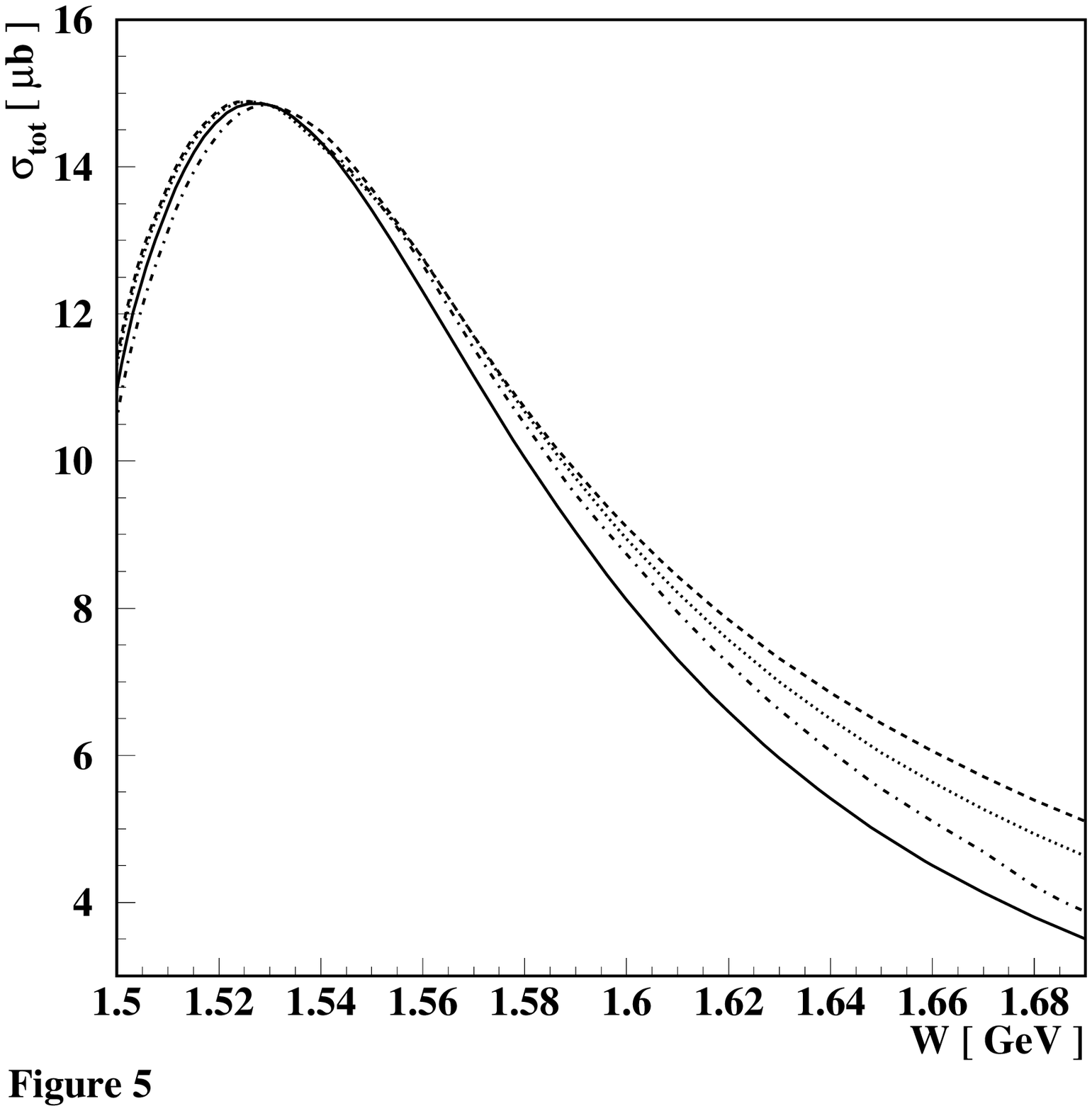}}
\label{Fig.5}
\end{figure}

\end{document}